\pdfoutput=1
\documentclass[sigconf]{aamas}

\setcopyright{ifaamas}
\acmConference[AAMAS '24]{Proc.\@ of the 23rd International Conference
on Autonomous Agents and Multiagent Systems (AAMAS 2024)}{May 6 -- 10, 2024}
{Auckland, New Zealand}{N.~Alechina, V.~Dignum, M.~Dastani, J.S.~Sichman (eds.)}
\copyrightyear{2024}
\acmYear{2024}
\acmDOI{}
\acmPrice{}
\acmISBN{}

\usepackage{booktabs}

\usepackage{standalone}

\usepackage{multirow}
\newcommand{\diagbox}[2]{#1: #2}

\usepackage{subfig}
\usepackage{xcolor}

\usepackage{pifont}

\usepackage{xspace}

\newcommand{\framework}{\fsl{Nest}\xspace}
\newcommand{\PrimitiveSoc}{\fsc{Primitive}\xspace}
\newcommand{\PrimitiveSocBrief}{\fsc{Prim.}\xspace}
\newcommand{\SanctioningSoc}{\fsc{Penalty}\xspace}
\newcommand{\SanctioningSocBrief}{\fsc{Pen.}\xspace}
\newcommand{\EmoteSoc}{\fsc{Emote}\xspace}
\newcommand{\TellSoc}{\fsc{Tell}\xspace}

\definecolor{revision}{rgb}{0.29, 0.0, 0.51}
\newcommand{\revisedst}[1]{}
\newcommand{\revised}[1]{#1}

\usepackage[normalem]{ulem}

\newcommand{\mypara}[1]{\paragraph{#1}}

\usepackage{siunitx}
\newcommand{\nint}[1]{\num[round-precision=0]{#1}}

\newcommand{\ntwo}[1]{\num[round-precision=2]{#1}}
\newcommand{\nthree}[1]{\num[round-precision=3]{#1}}
\newcommand{\nlow}{$<$\nthree{0.001}}

\usepackage{listings}
\definecolor{codegreen}{rgb}{0,0.6,0}
\definecolor{codegray}{rgb}{0.5,0.5,0.5}
\definecolor{codepurple}{rgb}{0.58,0,0.82}
\definecolor{backcolour}{rgb}{0.95,0.95,0.92}

\lstdefinestyle{mystyle}{
    backgroundcolor=\color{backcolour},
    commentstyle=\color{codegreen},
    keywordstyle=\color{magenta},
    numberstyle=\tiny\color{codegray},
    stringstyle=\color{codepurple},
    basicstyle=\ttfamily\small,
    breakatwhitespace=false,
    breaklines=true,
    captionpos=b,
    keepspaces=true,
    numbers=left,
    numbersep=5pt,
    showspaces=false,
    showstringspaces=false,
    showtabs=false,
    tabsize=2
}

\newtheorem{example}{Example}

\newcommand{\fsc}{\textsc}
\newcommand{\fsl}{\textsl}

\DeclareMathAlphabet{\mathsl}{OT1}{ptm}{m}{sl}

\RequirePackage[np]{numprint}
\npdecimalsign{.}

\newcommand{\fsub}[1]{\texorpdfstring{\textsubscript{#1}}{#1}}

\newcommand{\shyphen}{\text{--}} 

\definecolor{mpsNavy}{rgb}{0.01,0.01,0.5}
\definecolor{mpsGray}{rgb}{0.85,0.84,0.84}
\definecolor{lightGray}{rgb}{0.95,0.94,0.94}

\usepackage{tikz}
\usetikzlibrary{arrows, matrix, positioning, shapes}

\tikzstyle{strong}=[thick,->,>=stealth']

\usepackage{pgfplots}
\pgfplotsset{compat=newest}
\usetikzlibrary{pgfplots.statistics}
\usepgfplotslibrary{groupplots}
\usepgfplotslibrary{statistics}

\newcommand{\mpsHide}[1]{}

\title{\texorpdfstring{Norm Enforcement with a Soft Touch:\\ Faster Emergence, Happier Agents}{Norm Enforcement with a Soft Touch: Faster Emergence, Happier Agents}}

\acmSubmissionID{414}



\author{Sz-Ting Tzeng}
\affiliation{%
  \institution{North Carolina State University}
  \city{Raleigh}
  \state{NC}
  \country{US}
  \postcode{27695}
}
\email{stzeng@ncsu.edu}

\author{Nirav Ajmeri}
\affiliation{%
  \institution{University of Bristol}
  \city{Bristol}
  \state{}
  \country{UK}
  \postcode{BS8 1UB}
}
\email{nirav.ajmeri@bristol.ac.uk}

\author{Munindar P. Singh}
\affiliation{%
  \institution{North Carolina State University}
  \city{Raleigh}
  \state{NC}
  \country{US}
  \postcode{27695}
}
\email{mpsingh@ncsu.edu}

\begin{abstract}
A multiagent system is a society of autonomous agents whose interactions can be regulated via social norms.
In general, the norms of a society are not hardcoded but emerge from the agents' interactions.
Specifically, how the agents in a society react to each other's behavior and respond to the reactions of others determines which norms emerge in the society.
We think of these reactions by an agent to the satisfactory or unsatisfactory behaviors of another agent as communications from the first agent to the second agent.
Understanding these communications is a kind of social intelligence: these communications provide natural drivers for norm emergence by pushing agents toward certain behaviors, which can become established as norms.
Whereas it is well-known that sanctioning can lead to the emergence of norms, we posit that a broader kind of social intelligence can prove more effective in promoting cooperation in a multiagent system.

Accordingly, we develop \emph{\framework}, a framework that models social intelligence via a wider variety of communications and understanding of them than in previous work.
To evaluate \framework, we develop a simulated pandemic environment and conduct simulation experiments to compare \framework with baselines considering a combination of three kinds of social communication: \emph{sanction}, \emph{tell}, and \emph{hint}.

We find that societies formed of \framework agents achieve norms faster.
Moreover, \framework agents effectively avoid undesirable consequences, which are negative sanctions and deviation from goals, and yield higher satisfaction for themselves than baseline agents despite requiring only an equivalent amount of information.
\end{abstract}

\begin{document}

\maketitle

\section{Introduction}
\label{sec:introduction}
Social norms characterize collective and acceptable group conduct and regulate agent behavior.
Norms may be imposed top-down (as legal norms are) \citep{IJCAI-23:deviation} or emerge bottom-up (when agents learn acceptable behaviors from each other) \cite{Savarimuthu2011norm}.
Our interest is in the emergence of norms bottom-up while accommodating top-down norms.
A norm emerges in a society when a substantial majority of its agents take the same action in similar circumstances \cite{Morris2019norm,Savarimuthu2011norm}.

We posit that the emergence of norms is driven by three kinds of social communication from one agent to another in response to the first agent observing certain behaviors by the second agent in certain situations. These kinds of communication are
\begin{itemize}
\item \emph{Sanction:} a punishment or reward \cite{Nardin2016sanction} for an observed behavior,
\item \emph{Tell:} a direct normative message or explicit communication of approval or disapproval \cite{Andrighetto2013punish} of an observed behavior, and
\item \emph{Hint:} an implicit communication conveying a positive or a negative attitude toward an observed behavior.
\end{itemize}

\begin{example}{\textbf{Sanction.}}
\label{ex:punishment}
Becka is symptomatic with COVID-19.
Alice meets Becka in a cafe and notices Becka's symptoms.
Alice reports the violation of the healthcare guidance, leading to Becka being required to undergo a compulsory quarantine at a designated facility.
\end{example}

\begin{example}{\textbf{Tell.}}
\label{ex:normative_message}
Charlie notices Becka's suspicious symptoms and begins to worry about his safety.
He tells Becka that if she roams in public while symptomatic will lead her to be required to undergo a compulsory quarantine.
\end{example}

\begin{example}{\textbf{Hint.}}
\label{ex:hint}
David notices Becka's symptoms and expresses coldness near her.
Becka interprets David's negative attitude as a reaction to her apparent violating safety guidelines by sniffling near him.
Becka feels guilty for wandering out while being symptomatic and is anxious about the possibility of being reported.
\end{example}

The above social communications convey normative information from which Becka learns that her behavior was inappropriate.
In addition, third parties who observe Becka's behavior and follow the social communications directed at Becka may alter their behavior without directly having to be told.

Messages and hints drive subtle forms of social learning, as in human societies, but have not been adequately studied. A hint is a softer kind of social communication that has yet to be studied as a driver of norm emergence.
We investigate the following research question.
\mypara{RQ\fsub{social-communication}}
How does considering hints and normative messages in addition to sanctions influence norm emergence?

To address RQ\fsub{social-communication}, we define two expressions of normative information: explicit normative message \cite{Andrighetto2013punish} and implicit hint as information.

\mypara{Contributions}
We propose \emph{\framework} (for \fsl{Norm Enforcement with a Soft Touch}), a framework that accommodates norms imposed top-down and enables norm emergence.
\framework includes normative information from three types of social communication; that information facilitates social learning.

We evaluate \framework experimentally via a simulation of a pandemic scenario.
We examine societies characterized by three distinct social communication types: a sanction, a tell or direct messaging, and a hint.
Our results demonstrate that introducing normative information communicated via hints and direct messaging enables faster norm emergence, avoids undesirable consequences such as negative sanctions and deviation from goals, and yields higher satisfaction overall in a society.
\revised{Notably, in societies with low vaccination rates, individuals learn that engaging in self-isolation is praiseworthy, and short-term compromises prevent major penalties.}

\mypara{Organization}
Section~\ref{sec:framework} introduces key concepts of \framework and describes how agents' decision-making works in \framework.
Section~\ref{sec:simulation_environment} details the pandemic simulation we create to evaluate \framework.
Section~\ref{sec:results} presents results from our simulation experiments.
Section~\ref{sec:related-work} discusses other relevant research.
Section~\ref{sec:discussion} concludes with a summary of our findings, limitations, threats to validity, and future directions.

\section{\framework}
\label{sec:framework}
A \framework agent selects actions considering their goals, environmental norms, and social communication \cite{AAMAS-17:Arnor,Marsella2009ema,Savarimuthu2011norm,Singh1994MAS}.

A \framework agent learns from observations.
Following Example~\ref{ex:normative_message}, on receiving Charlie's message, Becka learns that she may be reported to local authorities if she does not self-isolate.
In Example~\ref{ex:hint}, Becka may have misread David's coldness as directed at her.

When making decisions, a \framework agent activates those norms related to itself based on its knowledge \cite{Argente2020normative}.
The normative reasoning process enables \framework agents to reason over the possible outcome of norm compliance or violation.
After executing a chosen action, the agent checks if a norm has been fulfilled or violated.
The compliance and violation of norms then triggers social communication.

\subsection{Key Concepts}
We now introduce the key concepts in \framework.

\paragraph{Goal} A goal is a condition over the state of the world that an agent wants to achieve.
The outcome of a goal upon performing a selected action is a binary value: achieved or not.

\paragraph{Norm} A norm defines the relationship between an agent (subject) on whom the norm is focused and an agent (object) with respect to whom the norm arises.
An agent can invest effort on its norm or has its freedom curtailed by the object.
\revised{We consider two norm types: Commitment and Prohibition \cite{TIST-13-Governance}.}
\revised{
A norm has two other elements: (1) antecedent, the conditions under which the norm is activated, and (2) consequent, the conditions under which the norm is satisfied or violated.}
A commitment norm can either be satisfied or violated when the consequent holds or not, respectively. A prohibition norm is violated when the consequent holds and is satisfied if the consequent never holds \citep{AAMAS-16:Custard}.

\paragraph{Sanction} A sanction refers to a positive, negative, or neutral reaction directed from one agent toward another. A sanction is typically in response to a norm satisfaction or norm violation.
\revised{Sanctions in the real world include subtle instances like the expression of emotions \cite{Nardin2016sanction}.}

\paragraph{Tell} A tell or a normative message specifies the cause and the effect.
The effect describes a potential reward or punishment.
For example, Charlie's specification in Example~\ref{ex:normative_message} includes whether a norm is satisfied or violated and why.

\paragraph{Hint}
A hint is an indirect clue that an agent expresses toward another agent to guide its behavior.
A hint requires the receiver to infer the intended meaning.
We model hints as subtle soft communications triggered by norm satisfaction or violation.

\paragraph{Reward Shaping}
This refers to supplemental or ``shaping'' rewards (in addition to those from the environment) provided to agents to move toward a certain goal or to encourage selecting a certain action in a certain set of states \cite{Marom2018belief}.
Here, we consider normative information from a tell or a hint as advice on potential soft sanctions with different probability. That is, communication types---\fsl{tell} and \fsl{hint}---are inferred as positive or negative rewards to encourage or discourage taking specific actions.

\subsection{Decision-Making}
\label{sec:decision-making}

An agent's behaviors include acting to maximize its payoff, and engaging in social communication.

\paragraph{Action selection}
An agent selects an action that satisfies its goal and maximizes its actual and possible payoff.
In the examples of Section~\ref{sec:introduction}, Becka decides whether to go to the cafe depending on her goals and her understanding of norms.

\paragraph{Social communication}
An agent observes other agents' behaviors and communicates via sanctions, messages, or hints if the behaviors conflict with norms.
In Example~\ref{ex:punishment}, Alice \revised{negatively} sanctions Becka based on the healthcare guidance of staying at home when symptomatic. In Example~\ref{ex:normative_message}, Charlie sends Becka a direct message. In Example~\ref{ex:hint}, David's coldness toward Becka shows his disapproval.

\paragraph{Reward Shaping}
A tell or a hint serves as a look-ahead advice on what will happen after a specific action.
A shaping reward can be defined as $r^\prime = r + F$ where $r$ is the original reward function, and $F$ is the shaping reward function.
With messages or hints, $F$ defines the difference of potential values. Here, $\Phi$ is a potential function that gives hints on states.
And, $\kappa$ defines the probability of a reward from the knowledge or information.

\begin{equation}
F(s, a, s^\prime, a^\prime) = \gamma\Phi(s^\prime, a^\prime)\kappa - \Phi(s, a)
\end{equation}

\section{Simulation}
\label{sec:simulation_environment}

We evaluate \framework via a simulated pandemic scenario where how agents behave influences the spread of a pandemic.
We built our pandemic environment using Mesa \cite{Masad2015mesa}, a Python-based simulation framework.
Our focus is not to model the realism of contagion spreading but to investigate how social communication influences norms.
Agents in the simulation use reinforcement learning to learn the relationship between objectives and normative behaviors.

\subsection{Pandemic Scenario}
\label{sec:environment}

In the simulated environment, an agent moves between four places: home (unique for each agent), park, cafe, and (vaccination) clinic.
An agent has a goal to {rest, hike, shop, be\_vaccinated}.
It selects actions from among the following: stay\_home, visit\_park, visit\_cafe, and visit\_clinic.
Any two agents present at the same place may interact with equal probability at each step.
Agents have the capacity to interpret each other's communications, and all forms of social communication are genuine and honest.

At each step, an agent observes its environment and moves based on factors such as
death, goal satisfaction, sanctions, messages, hints, and norm satisfaction or violation.
After all agents move, they evaluate each other's behaviors and communicate accordingly.
Communications directed to one agent are visible to other agents in the same location.
An agent who witnesses social communication between others can learn from the communication.

\subsection{Disease Model}
\label{sec:disease_model}

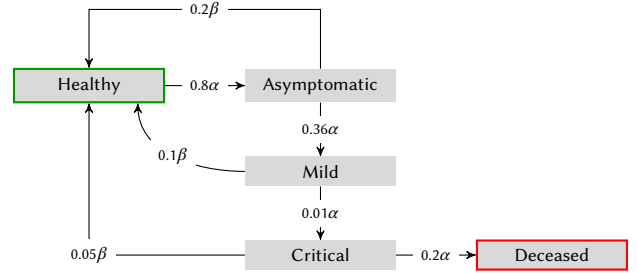
\begin{figure}
\centering
\begin{tikzpicture}
\usetikzlibrary{arrows}
\tikzstyle{every text node part/.style}=[align=center]
\tikzstyle{stage}=[font=\sffamily\footnotesize,fill=gray!30,text width=50,text centered]
\tikzstyle{move}=[->,>=stealth']
\tikzstyle{edgeLabel}=[font=\sffamily\scriptsize,fill=white]


\matrix[row sep=20,column sep=30] {
\node[stage,draw=green!60!black,thick] (Healthy) {Healthy};
&\node[stage] (Asymptomatic) {Asymptomatic};\\

&\node[stage] (Mild) {Mild};\\

&\node[stage] (Critical) {Critical};
&\node[stage,draw=red,thick] (Deceased) {Deceased};\\
};

\draw[move] (Healthy) -- node[edgeLabel] {0.8$\alpha$} (Asymptomatic);

\draw[move] (Asymptomatic) -- node[edgeLabel] {0.36$\alpha$} (Mild);
\draw[move] (Asymptomatic) -- +(0,1) -| node[edgeLabel,pos=0.25] {0.2$\beta$} (Healthy);

\draw[move] (Mild) -- node[edgeLabel] {0.01$\alpha$} (Critical);
\draw[move] (Mild) to[out=180,in=270] node[edgeLabel] {0.1$\beta$} (Healthy.340);

\draw[move] (Critical) -| node[edgeLabel] {0.05$\beta$} (Healthy);
\draw[move] (Critical) -- node[edgeLabel] {0.2$\alpha$} (Deceased);

\end{tikzpicture}
\caption{
Disease model with state transition probabilities.
The transition for healthy agents applies when coming in contact with those who are infected.
Here, $\alpha$ and $\beta$ parameterize the transition probabilities for vaccination and home rest, respectively.
In our study, $\alpha = 0.5$ means vaccinated, and $\alpha = 1.0$ means not vaccinated. And, $\beta = 2.0$ means at home and $\beta = 1.0$ means not at home.
For example, the edge weight from Mild to Critical can be read as \ntwo{0.01} for vaccinated agents and \num[round-precision=3]{0.005} for unvaccinated agents.
The probability of remaining in a state is $1 -$ the probability of evolving to the next state.
}
\label{fig:not-SEIRV}
\end{figure}

Our disease model is simplified from the Susceptible-Exposed-Infected-Recovered (SEIR) model
\cite{yang2020mathematical,annas2020stability} and captures the effectiveness of vaccines.
As shown in Figure~\ref{fig:not-SEIRV}, each agent begins in a healthy state.
Upon encountering an infected agent (not shown), it transitions to the asymptomatic phase of the disease, showing no symptoms despite being infected.
As the symptoms progress, an agent becomes mildly symptomatic, then critically symptomatic, and in the worst-case, deceased.
Vaccination offers protection by reducing the probability that agents can become infected and advance toward critical symptomatic or deceased.
Home-based recovery is the primary treatment during the pandemic.

We base the probabilities of how COVID-19 evolves on \citet{poletti2020probability}.
We set the infection probability to 80\% and the effectiveness of vaccination at 50\% to represent a more infectious variant and speed up the simulation.
Apart from vaccination, we set the probability of the symptoms to progress as Figure~\ref{fig:not-SEIRV}.
The intuition is that each infected person provides an opportunity for the symptoms to progress to the next phase or recover.

We assume home stay improves recovery from infection.
Below, we write ``isolation'' when a home stay is voluntary and ``quarantine'' when it is forced.
We set a 50\% probability to \revised{negatively} sanction those who exhibit mild symptoms and an 80\% probability to sanction those who exhibit critical symptoms but are not isolated.
Table~\ref{tbl:uncertainty} shows partial observability of the health states of others (e.g., a healthy person who has watery eyes because of an accident with pepper may be perceived as sick (Mild)).

\begin{table}[!htb]
\centering
\caption{Imperfect observation of another's health state.
}
\label{tbl:uncertainty}
\begin{tabular}{l r r r r}\toprule
\diagbox{Actual}{Belief}  & Healthy  & Mild & Critical \\\midrule
Healthy      & \ntwo{0.8} & \ntwo{0.1} & \ntwo{0.1} \\
Asymptomatic & \ntwo{0.5} & \ntwo{0.5} & \ntwo{0.0} \\
Mild         & \ntwo{0.3} & \ntwo{0.6} & \ntwo{0.1} \\
Critical     & \ntwo{0.1} & \ntwo{0.3} & \ntwo{0.6} \\

\bottomrule
\end{tabular}
\end{table}

\subsection{Social Norm}
\label{sec:existing_social_norm}

We initialize the environment with a social norm that healthy agents prohibit infected agents from staying in a public space.
We frame the norm as a prohibition as below.

\begin{lstlisting}
norm type   = {Prohibition},
subject     = {Infected_Agent},
object      = {Healthy_Agent},
antecedent  = {obs_health=[MILD, CRITICAL]},
consequent  = {loc=[PARK, CAFE, CLINIC]}
\end{lstlisting}
When the antecedent and consequent both hold, the prohibition is violated, and a sanction is given to the subject.
The sanction is a numerical reward, in this case negative, to the agent who violates the prohibition.
A negative reward represents an undesirable state. 
When an agent receives normative information from others indicating or stating this prohibition, it considers the sanction a potential reward $\Phi$.
According to the social communication type, the agent infers 
the shaping reward.

\subsection{Normative Information Communication}
\label{sec:normative_info_communication}

We formalize normative information via conditionals indicating that the stated consequent will be brought out when the presented antecedent holds.
For example:
\begin{lstlisting}
sender     = {Observer_Agent},
receiver   = {Actor_Agent},
info type  = {MESSAGE},
antecedent = {obs_health=CRITICAL,loc=CAFE},
consequent = {PUNISHMENT}
\end{lstlisting}

\subsection{Types of Societies}
\label{sec:agent_type}

We consider societies based on three kinds of social communication: sanctioning, telling or direct messaging, and providing hints.

\subsubsection{Baseline 1: \PrimitiveSoc society}
Agents apply no social communication and act solely based on goal satisfaction (payoffs).

\subsubsection{Baseline 2: \SanctioningSoc society}
Agents obtain negative sanctions for violating a social norm, as in Example~\ref{ex:punishment}.
Healthy agents may punish infected agents who enter a public space by forcing them to quarantine.
In that case, within the following few timesteps, a punished agent's position is changed to home regardless of its wishes.

\subsubsection{Baseline 3: \EmoteSoc society}
This society is a variant of \framework without shaping rewards.
With some social norm in mind, agents who violate or satisfy the social norm receive a social communication, such as expressed emotions, from others.
These agents may experience guilt or pleasure based on norm violation or satisfaction.
\EmoteSoc is adapted from \citet{Tzeng2021noe}.
Infected agents who wander in a public space may receive expressed emotions and feel bad about their norm violation and may be forced to quarantine at home by healthy agents.

\subsubsection{Baseline 4: \TellSoc society}
This society is a variant of \framework without the \fsl{hint} part, as in
Example~\ref{ex:normative_message}.
Agents learn social norms and convey normative messages upon witnessing a norm violation.
The normative message is adapted from \citet{Andrighetto2013punish} and includes what sanctions an agent will receive if it violates a norm.
Healthy agents interacting with infected agents in a public space convey the social norm of staying away from public spaces to the infected agents.
Also, infected agents in a public space may be forced to quarantine by healthy agents.

\subsubsection{\framework: Hint Society}
A society with our proposed agents.
\framework agents learn norms from sanctions and hints.
In the simulation, infected agents who wander in public spaces receive hints from healthy agents via which they infer the normative information about staying away from healthy agents.
This information from hints provides shaping rewards to agents to learn norms.
\framework agents may also experience pleasure or guilt based on norm violation or satisfaction.
In addition, as in \SanctioningSoc, \EmoteSoc, and \TellSoc, infected agents in a public space may be forced to quarantine at home by healthy agents.

\subsection{Metrics}
\label{sec:metrics}
We compute six measures to evaluate \framework.
M\fsub{Healthy}, M\fsub{Infected}, M\fsub{Deceased}, M\fsub{Infections}, M\fsub{Vaccinated}, and M\fsub{Goal} help identify the consequences of agents' behaviors or norm emergence.
Moreover, these measures provide insights into why specific norms emerge.
M\fsub{Home} and M\fsub{Quarantine} yield the percentage of self-isolation behaviors.

\begin{description}
\item[M\fsub{Healthy}] Percentage of agents who are healthy.

\item[M\fsub{Infected}] Percentage of agents who are infected.

\item[M\fsub{Deceased}] Percentage of agents who are deceased.

\item[M\fsub{Infections}] Average number of infections.

\item[M\fsub{Vaccinated}] Percentage of agents who are vaccinated.

\item[M\fsub{Home}] Percentage of infected agents who stay home.

\item[M\fsub{Quarantine}] Number of agents forced to quarantine at home. This measure maps to the social communication type sanctioning.

\item[M\fsub{Goal}] The average goal satisfaction among agents.
\end{description}

A norm emerges when the proportion of agents following the same behavior exceeds a threshold. We consider 90\% as the threshold \cite{Delgado2002emergence}.

\subsection{Hypotheses}
\label{sec:hypotheses}

We evaluate three hypotheses.
For each of them, we test statistical significance with respect to its null hypothesis via the independent t-test.
We adopt Glass' $\Delta$ as a measure of effect size since the standard deviations are different between the societies \cite{Glass1976primary,Grissom2012effect}.
In addition, we adopt Cohen's \cite{Cohen1988statistics} descriptors to interpret the effect size.
Specifically, an effect size magnitude in [0,0.2) indicates that the difference is negligible and magnitudes in [0.2--0.5), [0.5--0.8), and $[0.8,\infty)$ respectively indicate a small, a medium, and a large effect.

\mypara{H\fsub{Disease control}} Societies considering hints have better control over disease spread than the societies that do not consider hints.
We compare societies with respect to M\fsub{Healthy}, M\fsub{Infected}, M\fsub{Deceased}, M\fsub{Infections}, and M\fsub{Vaccination}.

\mypara{H\fsub{Isolation}} Societies considering hints yield improved isolation than other societies.
We compare societies with respect to M\fsub{Home}, M\fsub{Quarantine}, and M\fsub{Infected}.

\mypara{H\fsub{Goal}} Agents in \framework have higher goal satisfaction than other societies.
We compare societies with respect to M\fsub{Goal}.

\subsection{Experimental Setup}
\label{sec:experiment-setup}

Table~\ref{tbl:general_payoffs}
lists the elements of reward function for an agent, including extrinsic rewards from the environment and intrinsic rewards from an agent's internal state.
For data efficiency, we apply policy parameter sharing \cite{gupta2017cooperative} based on the assumption of the bystander.
We consider \nint{100} agents (\nint{30} are infected initially) with the simulated world lasting for \nint{2000} steps.
Each society stabilizes within \nint{1500} steps.
We train our agents for \nint{100000}
steps, and report results averaged over 20 runs.

We consider normative information as shaping rewards that are part of intrinsic rewards.
Specifically, we incorporate knowledge of being punished in the future from tell or hint into our simulation.
Table~\ref{tbl:general_payoffs} lists elements based on \framework agents appraises their states.

\begin{table}[!htb]
\centering
\caption{Reward function.
Sanctioning means quarantine.
An agent's extrinsic rewards come from the environment and intrinsic rewards come from its current internal state.
Norm satisfaction or violation is based on the action and the perceived health state of others instead of the actual health state.
}
\label{tbl:general_payoffs}
\begin{tabular}{l l n{5}{1}}\toprule
Component & Type & {Reward} \\\midrule
Deceased & Extrinsic & -2 \\
Sanctioning & Extrinsic & -1 \\
Goal satisfaction & Intrinsic & +1\ \\
Goal violation & Intrinsic & -1\ \\
Norm satisfaction (self) & Intrinsic & +0.5\ \\
Norm violation (self) & Intrinsic & -0.5\ \\
Norm satisfaction (other) & Extrinsic & +0.5\ \\
Norm violation (other) & Extrinsic & -0.5\ \\
\bottomrule
\end{tabular}
\end{table}

\subsubsection{Information Balance}
\label{sec:information-balance}

As messages and hints provide additional normative information to learn, we keep the social communication distributions at the same level to balance the amount of information that agents receive from a combination of communication types.
Specifically, since enhancing communication can improve learning, we adjust the communication-type distributions to balance the information across the societies.
Table~\ref{tbl:communication_prob} lists the probability distribution over the various types of social communication we apply for each society.
Specifically, we have the following.

\begin{itemize}

\item \fsl{Sanction} is the probability of an agent being compelled to quarantine.
An agent considers a negative sanction as a punishment.

\item \fsl{Tell} is the probability that an agent receives a \fsl{tell} from its neighbors.
An agent regards a \fsl{tell} as a potential sanction: thus it carries 50\% of the weight of an actual sanction.

\item \fsl{Emote} is the probability that an agent receives an implicit communication such as via an expressed emotion conveying a positive or a negative attitude. These communications serve as subtle sanctions but are distinguished from ``sanctions'' as formalized here.

\item  \fsl{Hint} is the probability that an agent receives a positive or negative attitude as an implicit communication.
The agent interprets a hint as a potential sanction
with \revised{30\% of the weight of the actual sanction.}
\end{itemize}

\begin{table}[!htb]
\centering
\caption{Social communication distributions. In each society, agents send a combination of the three kinds of social communication.
To balance the amount of information, we adjust the distributions immediate (I) and potential (P) rewards as here.
Here, wI and wP describe the weights associated with the rewards.
\EmoteSoc expresses other-directed attitudes towards others' behaviors as sanctions and has a self-directed attitude toward the self.
}
\label{tbl:communication_prob}
\begin{tabular}{@{}l@{~~} r r r r r r @{~~~} r@{}}
\toprule
 & \multicolumn{5}{c}{Social communication distribution} &  & \\\cmidrule{2-6}
 \multirow{2}{*}{\parbox{1.5cm}{Society}} & Sanction & Tell & Emote & Hint & None & \multirow{2}{*}{wI} & \multirow{2}{*}{wP}\\
 & I & P & I & I+P & & \\\midrule
\PrimitiveSoc     & 0\% & 0\% & 0\% & 0\% & 100\% & 0 & 0 \\
\SanctioningSoc   & 38\%  & 0\% & 0\% & 0\% & 62\% & 1 & 0 \\
\TellSoc       & 20\% & 18\% & 0\% & 0\% & 62\% & 1 & 0.5 \\
\EmoteSoc      & 20\% & 0\% & 18\% & 0\% & 62\% & $1+0.5$ & 0 \\
\framework: Hint        & 20\% & 0\% & 0\% & 18\% & 62\% & $1+0.5$ & 0.3\\


\bottomrule
\end{tabular}
\end{table}

\subsubsection{Reinforcement Learning Parameters for Social Communication}
\label{sec:rl}

\framework and baseline agents employ Q-learning \cite{Watkins1992q} to learn norms.
Q-learning is a model-free reinforcement learning algorithm that learns by computing the action-state value $Q(s, a)$ (Q value), which indicates the expected and cumulative rewards for each state and action.
The Q function computes Q values with the weighted average of the old value and the new information, and is given by:

\begin{equation}
Q(s, a)=Q(s, a)+\alpha \times
(r_t+\gamma \max \limits_{a^\prime} Q(s^\prime, a^\prime)-Q(s, a))
\end{equation}
where $Q(s, a)$ is the expected value for performing action $a$ in state $s$.
Here, $\alpha$ is the learning rate and $\gamma$ is the reward discount rate.
$s^\prime$ refers to the next state and $a^\prime$ refers to possible actions in $s^\prime$.
Q-learning finds the optimal policy by approximating the value of an action for a given state.

Messages give precise causality between claimed behaviors and possible sanctions.
On the contrary, hints provide subtle normative information for behaviors, which requires further inference.
While hints and messages provide different probability of possible sanctions, we model normative information with approving and disapproving attitudes as shaping rewards.
Specifically, we associate approving and disapproving attitudes with norm satisfaction and violation.
The potential rewards are calculated from the social communication based on the communication type.
We set $\kappa$ as \ntwo{0.3} for hints and $\kappa$ as \ntwo{0.5} for tells, where $\kappa$ is the probability of a possible sanction from normative information.
These parameters are selected empirically.
Note that the trends in results are consistent with changes in parameters.

\section{Experimental Results}
\label{sec:results}

We now discuss the results for RQ\fsub{social-communication}.
Table~\ref{tab:results_analysis_info}
summarizes the simulation results and the corresponding statistical analysis for RQ\fsub{social-communication}.
For each hypothesis, the metric reported is upon convergence.

\begin{table}[htb] 
\caption{
Comparing \framework with baseline societies on various metrics and their statistical analysis with Glass' $\Delta$. All p-values are {\nlow}.
    The metrics reported are calculated upon convergence.
    \framework defeats all the others on all the metrics. But \EmoteSoc is 3--3 with \TellSoc (2 ties). Surprisingly, \EmoteSoc also loses to \SanctioningSoc 3--5.
}

\label{tab:results_analysis_info}
\begin{tabular}{l l @{~~}S[table-format=2.2,round-precision=2] S[table-format=2.2,round-precision=2] S[table-format=2.2,round-precision=2] S[table-format=2.2,round-precision=2] S[table-format=2.2]}

        \toprule
        & & {\PrimitiveSocBrief} & {\SanctioningSocBrief} & {\EmoteSoc} & {\TellSoc} & {\framework} \\\midrule
        \multirow{10}{*}{\rotatebox[origin=c]{90}{H\fsub{Disease control}}}& M\fsub{Infected} & 13.2892 & 2.6480 & 3.7812 & 2.9587 & 0.2179  \\
        & $\Delta$ & 0.9578 & 0.2482 & 0.3044 & 0.2628 & $\shyphen$ \\\cmidrule{2-7}

        & M\fsub{Healthy} & 46.3058 & 77.5979 & 67.1149 & 76.2650 & 97.5444  \\
        & $\Delta$ & 17.5435 & 3.2194 & 4.7949 & 3.2314 & $\shyphen$ \\\cmidrule{2-7}

        & M\fsub{Deceased} & 41.0137 & 19.7541 & 29.1039 & 20.7762 & 2.0815  \\
        & $\Delta$ & 3.2529 & 5.7376 & 5.2656 & 5.4804 & $\shyphen$ \\\cmidrule{2-7}

        & M\fsub{Infections} & 48.3110 & 13.8329 & 19.0872 & 15.1617 & 2.0683 \\
        & $\Delta$ & 2.5926 & 6.2084 & 5.5565 & 6.0002 & $\shyphen$ \\\cmidrule{2-7}

        & M\fsub{Vaccinated} & 82.4112 & 36.7246 & 32.6922 & 35.3318 & 93.5654 \\
        & $\Delta$ & 1.0253 & 16.2111 & 15.8621 & 14.7792 & $\shyphen$ \\\midrule

        \multirow{4}{*}{\rotatebox[origin=c]{90}{H\fsub{Isolation}}}& M\fsub{Home} & 0.6092 & 0.9642 & 0.9471 & 0.9516 & 0.9948  \\
        & $\Delta$ & 1.7611 & 0.2937 & 0.3810 & 0.3487 & $\shyphen$ \\\cmidrule{2-7}

        & M\fsub{Quarantine} & $\shyphen$ & 0.0258 & 0.0195 & 0.0208 & 0 \\
        & $\Delta$ & $\shyphen$ & 0.2569 & 0.2796 & 0.2569 & $\shyphen$ \\\midrule

        \multirow{2}{*}{\rotatebox[origin=c]{90}{H\fsub{Goal}}}& M\fsub{Goal} & 0.1874 & 0.2617 & 0.2333 & 0.2595 & 0.3078  \\
        & $\Delta$ & 3.0410 & 3.0064 & 3.6658 & 3.0428 & $\shyphen$ \\\bottomrule
    \end{tabular}
\end{table}

\subsection{H\fsub{Disease control}}

\begin{figure}[!htb]
\centering

    \begin{tikzpicture}
        \begin{axis}[
        title={},
        height=0.7\columnwidth,
        width=0.99\columnwidth,
        xlabel={Time in steps},
        ylabel={Infected},
        xticklabels={,,200,,600,,1000,,1400},
        xmin=-50,xmax=1550,
        ymin=-5,ymax=70,
        legend  columns=2,
        legend style={fill=none,draw=none}
        ]
        \addplot +[mark size=2.5, each nth point={100}, densely dotted] table [x=steps, y=infected, col sep=comma]
        {data/Ness/primitive_rolling_50.csv};
        \addplot +[mark size=2.5, each nth point={100}, dashdotted] table [x=steps, y=infected, col sep=comma]
        {data/Ness/sanctioning_rolling_50.csv};
        \addplot +[mark size=2.5, each nth point={100}, densely dashed, mark=triangle*] table [x=steps, y=infected, col sep=comma]
        {data/Ness/feeling_probE0.5_rolling_50.csv};
        \addplot +[mark size=2.5, each nth point={100}, dashed] table [x=steps, y=infected, col sep=comma]
        {data/Ness/message_prob_rolling_50.csv};
        \addplot +[green!50!black, mark color=green, mark size=2.5, each nth point={100}] table [x=steps, y=infected, col sep=comma]
        {data/Ness/hint_probE0.5_rolling_50.csv};
        \legend{\small \PrimitiveSoc, \small \SanctioningSoc, \small \EmoteSoc, \small \TellSoc, \small \framework}
        \end{axis}
    \end{tikzpicture}
    %
    \newline\newline
    \begin{tikzpicture}
        \begin{axis}[
        title={},
        height=0.7\columnwidth,
        width=0.99\columnwidth,
        xlabel={Time in steps},
        ylabel={Deceased},
        xticklabels={,,200,,600,,1000,,1400},
        xmin=-50,xmax=1550,
        legend style={fill=none}
        ]
        \addplot +[mark size=2.5, each nth point={100}, densely dotted] table [x=steps, y=deceased, col sep=comma]
        {data/Ness/primitive_rolling_50.csv};
        \addplot +[mark size=2.5, each nth point={100}, dashdotted] table [x=steps, y=deceased, col sep=comma]
        {data/Ness/sanctioning_rolling_50.csv};
        \addplot +[mark size=2.5, each nth point={100}, densely dashed, mark=triangle*] table [x=steps, y=deceased, col sep=comma]
        {data/Ness/feeling_probE0.5_rolling_50.csv};
        \addplot +[mark size=2.5, each nth point={100}, dashed] table [x=steps, y=deceased, col sep=comma]
        {data/Ness/message_prob_rolling_50.csv};
        \addplot +[green!50!black, mark color=green, mark size=2.5, each nth point={100}] table [x=steps, y=deceased, col sep=comma]
        {data/Ness/hint_probE0.5_rolling_50.csv};
        \end{axis}
    \end{tikzpicture}
    \newline\newline
    \begin{tikzpicture}
        \begin{axis}[
        title={},
        height=0.7\columnwidth,
        width=0.99\columnwidth,
        xlabel={Time in steps},
        ylabel={Vaccinated},
        xticklabels={,,200,,600,,1000,,1400},
        xmin=-50,xmax=1550,
        legend style={fill=none}
        ]
        \addplot +[mark size=2.5, each nth point={100}, densely dotted] table [x=steps, y=vaccinated, col sep=comma]
        {data/Ness/primitive_rolling_50.csv};
        \addplot +[mark size=2.5, each nth point={100}, dashdotted] table [x=steps, y=vaccinated, col sep=comma]
        {data/Ness/sanctioning_rolling_50.csv};
        \addplot +[mark size=2.5, each nth point={100}, densely dashed, mark=triangle*] table [x=steps, y=vaccinated, col sep=comma]
        {data/Ness/feeling_probE0.5_rolling_50.csv};
        \addplot +[mark size=2.5, each nth point={100}, dashed] table [x=steps, y=vaccinated, col sep=comma]
        {data/Ness/message_prob_rolling_50.csv};
        \addplot +[green!50!black, mark color=green, mark size=2.5, each nth point={100}] table [x=steps, y=vaccinated, col sep=comma]
        {data/Ness/hint_probE0.5_rolling_50.csv};
        \legend{}
        \end{axis}
    \end{tikzpicture}
\caption{
\framework results in the least infected and deceased agents with the highest vaccination rate among all societies.
However, despite a lower fraction of vaccinated agents, \EmoteSoc has fewer infected and deceased agents, and more healthy agents than other baselines.
The effect is large for the comparisons of M\fsub{Deceased} and M\fsub{Infections}\mpsHide{ with $p<0.001$}.
For M\fsub{Infected}, the effect is negligible for \EmoteSoc and small for \SanctioningSoc and \TellSoc and large for \PrimitiveSoc.
Appendix~\ref{sec:additional-results}
includes plots for the first 500 steps where the differences are noticeable.
}
\label{fig:seirv_results}
\end{figure}
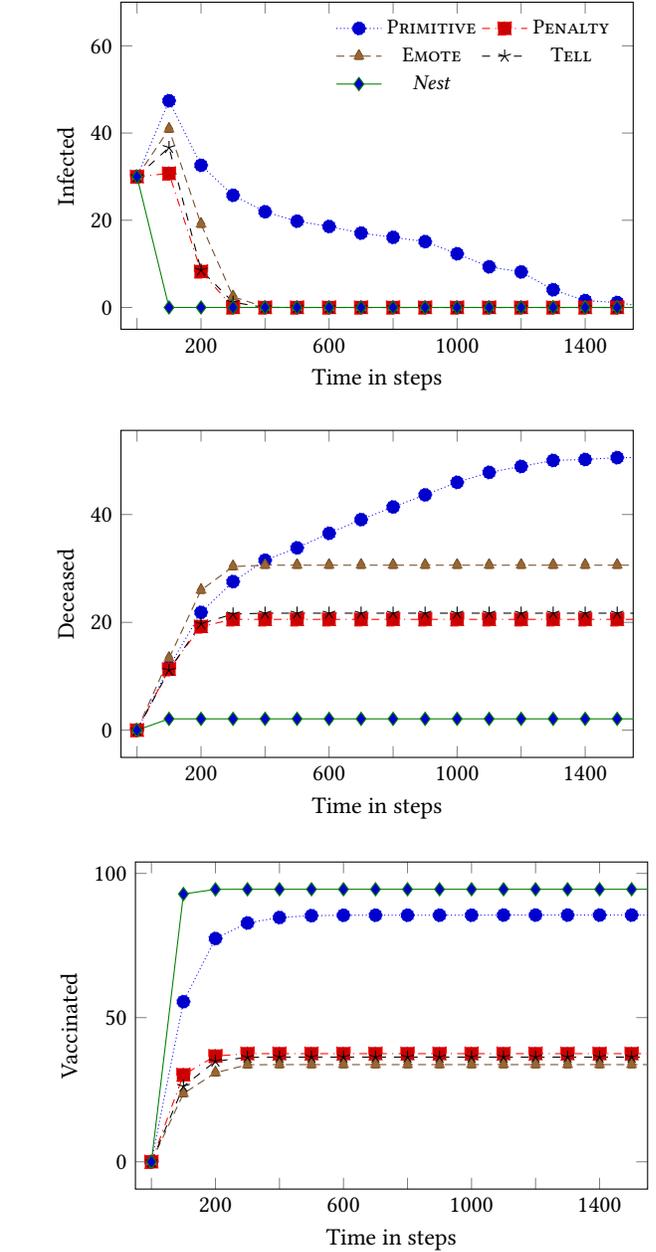

To evaluate H\fsub{Disease control}, we measure the proportion of healthy (M\fsub{Healthy}), infectious (M\fsub{Infected}), and deceased (M\fsub{Deceased}) agents.
We also track the average number of infections (M\fsub{Infections}) and vaccination rate (M\fsub{Vaccinated}).
Infectious agents include those who are asymptomatic, mild symptomatic, and critical.
Figure~\ref{fig:seirv_results} reports these metrics.
These simulations start from a 30\% infection rate in each society.
First, \framework has a lower fraction of infected agents (\num{0.21786606696651675}) than \PrimitiveSoc (\num{13.289235771893281}), \SanctioningSoc (\num{2.647976011994003}), \EmoteSoc (\num{3.781234382808596}), and \TellSoc (\num{2.9587456271864068}).
The effect is large for \PrimitiveSoc and small for \SanctioningSoc and \TellSoc, but negligible for \EmoteSoc\mpsHide{ with $p<0.001$}.

Second, \framework produces a larger number of  healthy agents (\num{97.54437781109446}) than \PrimitiveSoc (\num{46.30581616303146}), \SanctioningSoc (\num{77.59787606196902}), \EmoteSoc (\num{67.1148675662169}), and \TellSoc (\num{76.26504247876062}).
The effect is large\mpsHide{ with $p<0.001$}.

Third, \framework has a lower fraction of deceased agents (\num{2.081459270364818}) than \PrimitiveSoc (\num{41.013743128435785}), \SanctioningSoc (\num{19.754147926036982}), \EmoteSoc (\num{29.103898050974514}), and \TellSoc (\num{20.776211894052974}).
\framework has a lower M\fsub{Infections}(\num{2.0683283358320836}) than \PrimitiveSoc (\num{48.31099225387307}), \SanctioningSoc (\num{13.83286731634183}), \EmoteSoc (\num{19.087244377811096}), and \TellSoc (\num{15.16171364317841}).
The effect is large for each case\mpsHide{ and $p<0.001$ for the differences in the results}.

With regard to M\fsub{Vaccinated}, \framework has a higher vaccination rate (\num{93.56536731634183}) than \PrimitiveSoc (\num{82.41125475001917}), \SanctioningSoc (\num{36.72463768115942}), \EmoteSoc (\num{32.69220389805098}), and \TellSoc (\num{35.33178410794603}).
The effect is large\mpsHide{ and $p<0.001$}.

{With M\fsub{Vaccinated}, we observe that a vaccination norm emerges with a majority above 90\% in \framework.
Specifically, even without a top-down imposed shared expectation on vaccination, most agents in \framework learn that vaccination maximizes their payoff.}
\revised{The emerged vaccination norm can be articulated as below.}
\begin{lstlisting}
norm type   = {Commitment},
subject     = {Alive_Agent},
object      = {Other_Agent},
antecedent  = {vaccinated=FALSE;
               actual_health=[HEALTHY,
               ASYMPTOMATIC, MILD, CRITICAL]},
consequent  = {loc=[CLINIC]}
\end{lstlisting}
In societies where vaccination rates are low, agents learn that practicing self-isolation is commendable and that making short-term compromise
can help avoid major penalties.

\subsection{H\fsub{Isolation}}
\begin{figure}[!htb]
\centering
    \begin{tikzpicture}
    \begin{axis}[
    title={},
    height=0.7\columnwidth,
    width=0.99\columnwidth,
    xlabel={Time in steps},
    ylabel={Quarantine},
    xticklabels={,,200,,600,,1000,,1400},
    xmin=-50,xmax=1550,
    legend  columns=2,
    legend style={fill=none,draw=none}
    ]
    \addplot +[mark size=2.5, each nth point={100}, densely dotted] table [x=steps, y=forced_quarantine, col sep=comma]
    {data/Ness/primitive_rolling_50.csv};
    \addplot +[mark size=2.5, each nth point={100}, dashdotted] table [x=steps, y=forced_quarantine, col sep=comma]
    {data/Ness/sanctioning_rolling_50.csv};
    \addplot +[mark size=2.5, each nth point={100}, densely dashed, mark=triangle*] table [x=steps, y=forced_quarantine, col sep=comma]
    {data/Ness/feeling_probE0.5_rolling_50.csv};
    \addplot +[mark size=2.5, each nth point={100}, dashed] table [x=steps, y=forced_quarantine, col sep=comma]
    {data/Ness/message_prob_rolling_50.csv};
    \addplot +[green!50!black, mark color=green, mark size=2.5, each nth point={100}] table [x=steps, y=forced_quarantine, col sep=comma]
    {data/Ness/hint_probE0.5_rolling_50.csv};
    \legend{\small \PrimitiveSoc, \small \SanctioningSoc, \small \EmoteSoc, \small \TellSoc, \small \framework}
    \end{axis}
    \end{tikzpicture}
    \newline\newline
    \begin{tikzpicture}
    \begin{axis}[
    title={},
    height=0.7\columnwidth,
    width=0.99\columnwidth,
    xlabel={Time in steps},
    ylabel={Home},
    xticklabels={,,200,,600,,1000,,1400},
    xmin=-50,xmax=1550,
    legend style={at={(1.35,0.65)}, anchor=east},
    legend style={fill=none}
    ]
    \addplot +[mark size=2.5, each nth point={100}, densely dotted] table [x=steps, y=isolation, col sep=comma]
    {data/Ness/primitive_rolling_50.csv};
    \addplot +[mark size=2.5, each nth point={100}, dashdotted] table [x=steps, y=isolation, col sep=comma]
    {data/Ness/sanctioning_rolling_50.csv};
    \addplot +[mark size=2.5, each nth point={100}, densely dashed, mark=triangle*] table [x=steps, y=isolation, col sep=comma]
    {data/Ness/feeling_prob_rolling_50.csv};
    \addplot +[mark size=2.5, each nth point={100}, dashed] table [x=steps, y=isolation, col sep=comma]
    {data/Ness/message_prob_rolling_50.csv};
    \addplot +[green!50!black, mark color=green, mark size=2.5, each nth point={100}] table [x=steps, y=isolation, col sep=comma]
    {data/Ness/hint_prob_rolling_50.csv};
    \end{axis}
    \end{tikzpicture}

\caption{
Isolation is higher in \EmoteSoc and \framework (effect is small) than in societies that lack hints.
\framework puts fewer
agents in quarantine to achieve stable cooperation than \SanctioningSoc and \TellSoc.
The effect is negligible for \EmoteSoc and small for \SanctioningSoc and \TellSoc.
}
\label{fig:compliance}
\end{figure}
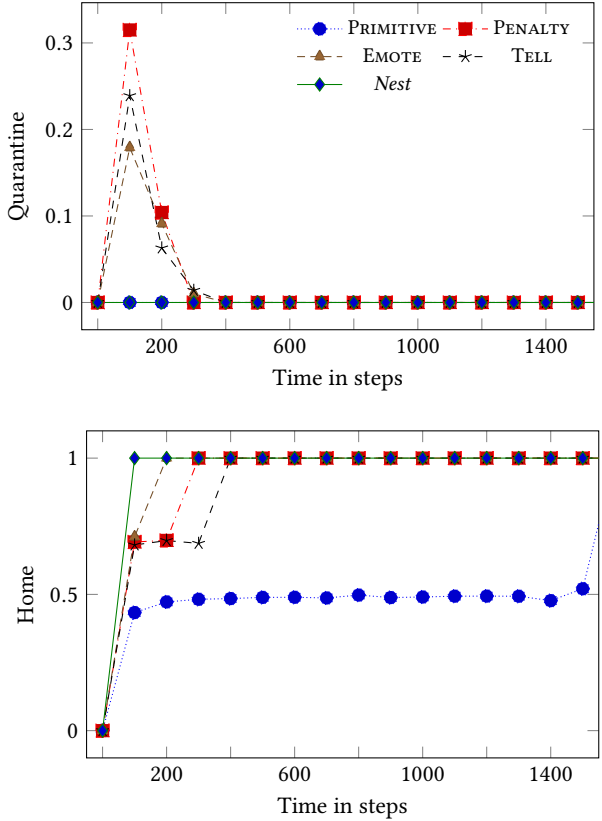

To evaluate H\fsub{Isolation}, we measure the proportion of infected agents who stay at home
(M\fsub{Home}), the number of agents in quarantine (M\fsub{Quarantine}), and the percentage of infected agents (M\fsub{Infected} in societies.
Figure~\ref{fig:compliance} exhibits plots comparing M\fsub{Home} and M\fsub{Quarantine}.

We observe that \framework yields a higher tendency to stay isolated (\num{0.9948144943296364}) when infected than \PrimitiveSoc (\num{0.609202720330372}), \SanctioningSoc (\num{0.964151694281233}), \EmoteSoc (\num{0.9470512577815399}), and \TellSoc (\num{0.9515999366612831}).
The effect is large for \PrimitiveSoc and small for the others\mpsHide{ with $p<0.001$}.

\framework has a lower M\fsub{Quarantine}
(\num{0})
than \SanctioningSoc (\num{0.025812093953023488}), \EmoteSoc (\num{0.019515242378810596}), and \TellSoc (\num{0.020839580209895053}).
The effect is small\mpsHide{ with $p<0.001$}.

From M\fsub{Home} and M\fsub{Quarantine} and M\fsub{Infected}, we observe that a norm emerges with a majority above 90\% in all societies other than \PrimitiveSoc.
\revised{The emerged self-isolation norm is as below.}
\begin{lstlisting}
norm type   = {Commitment},
subject     = {Infected_Agent},
object      = {Healthy_Agent},
antecedent  = {actual_health=[MILD, CRITICAL]},
consequent  = {loc=[HOME]}
\end{lstlisting}
Specifically, agents in societies with the mask-wearing
norm learn to comply with the norm and stay self-isolated when infected.
Furthermore, we see that this norm emerges fastest in \framework (\num{0.9948144943296364}).
With more subtle attitudes and information from hints, agents in \framework learn faster than those in \TellSoc.

\subsection{H\fsub{Goal}}

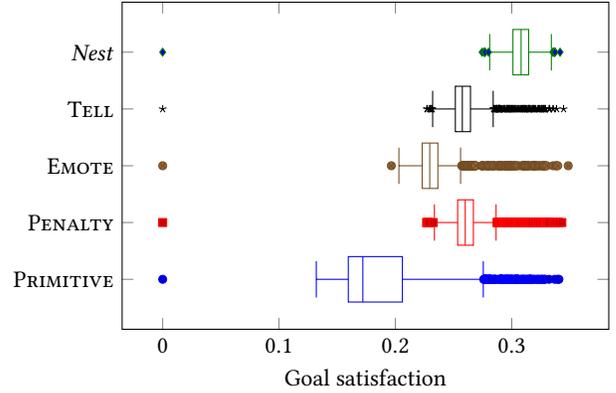
\begin{figure}[!htb]
\centering

    \begin{tikzpicture}
    \begin{axis}[
    title={},
    height=0.7\columnwidth,
    width=0.95\columnwidth,
    xlabel={Goal satisfaction},
    ytick={1,2,3,4,5},
    mark size=1.5,
    yticklabels={\PrimitiveSoc, \SanctioningSoc, \EmoteSoc, \TellSoc, \framework},
    scaled ticks=false, tick label style={/pgf/number format/fixed}
    ]
    \addplot+[boxplot]
 	table[y=desire_satisfaction, col sep=comma]
    {data/Ness/primitive.csv};
    \addplot+[boxplot]
 	table[y=desire_satisfaction, col sep=comma]
    {data/Ness/sanctioning.csv};
    \addplot+[boxplot]
 	table[y=desire_satisfaction, col sep=comma]
    {data/Ness/feeling_probE0.5.csv};
    \addplot+[boxplot]
 	table[y=desire_satisfaction, col sep=comma]
    {data/Ness/message_prob.csv};
    \addplot+[green!50!black,boxplot]
 	table[y=desire_satisfaction, col sep=comma]
    {data/Ness/hint_probE0.5.csv};
    \end{axis}
    \end{tikzpicture}

\caption{
\framework yields more goal satisfaction
than \PrimitiveSoc, \SanctioningSoc, \EmoteSoc, and \TellSoc.
The effect is small for
\EmoteSoc and large for the other societies\mpsHide{ with $p<0.001$}.
}
\label{fig:goal_satisfaction}
\end{figure}

To evaluate H\fsub{Goal}, we measure the goal satisfaction (M\fsub{Goal}) in societies.
Figure~\ref{fig:goal_satisfaction} plots M\fsub{Goal} in societies.
We observe that agents in \framework yield higher goal satisfaction (\num{0.3078}) than the \PrimitiveSoc (\num{0.18746475}), \SanctioningSoc (\num{0.26182125}), \EmoteSoc (\num{0.2333}), and \TellSoc (\num{0.25954197901049475}).
The effect is small 
for \EmoteSoc and large for the other societies\mpsHide{ with $p<0.001$}.

\section{Related Work}
\label{sec:related-work}

Research on norms and norm emergence closely relates to our contributions.
\citet{Andrighetto2013punish} show that a combination of verbal normative information, specifically positive normative content, and negative sanction
leads to higher and more stable cooperation---with both human subjects and agent-based simulation.
These models include normative reasoning but leave out softer notions of communication such as hints.
\citet{Kalia2019emotions} demonstrate how social communication such as emotions influences norm satisfaction.
Hints in \framework could be understood as emotions but hints serve both as a sanctioning approach and providing information about norms.

\citet{Bourgais2019ben} present an agent architecture that integrates cognition, personality, norms, and social relations with the conception of a contagion to simulate humans and improve the explainability of behaviors.
\citet{Argente2020normative} propose an abstract normative emotional agent architecture, an extension of BDI architecture that combines emotional, normative, and cognitive component.
\citet{Tzeng2021noe} combine normative model, a BDI model, and emotions for the decision-making process.
Agents in \framework learn from their interactions with the environment and further interpret norms from various kinds of social communication.

\citet{Mashayekhi2022Cha} propose a norm emergence framework that operates on conflict detection and inequity aversion.
Their framework enables agents to pass experience with utilities, associated states, and actions to agents of the same type.
\framework agents maximize personal payoff while social communications propel the norm emergence.

\citet{Dignum2020analysing} associate the interventions that governments can take and their economic and social consequences with the SEIR model since effective and sustainable solutions cannot exist without considering these factors.
\framework further takes social communication into consideration.

\citet{de2021quantifying} develop a large-scale data-driven agent-based simulation model where each agent reasons about their internal attitudes and external factors to simulate behavioral interventions in the real world.
\framework enables norm emergence and accommodates imposed norms.

\citet{DellAnna+20:sanctions} introduce a norm-revision component that uses data collected from interactions and an estimation of agents' preferences to modify sanctions at runtime.
 \citet{Lima2018gavel} enable agents to pick sanctions appropriate to the context.
\citet{Realpe2018balancing} present a model in which agents incorporate personal and normative considerations.
Specifically, agents make decisions that maximize their respective payoffs while appraising their group's social norms.
\framework defines the utility function based on
normative information learned from social communication.

\citet{Airiau2014emergence} propose a model that supports the emergence of social norms by learning from the interactions of a group of agents.
In \framework, we include cognition via social communication.
\citet{Hao2017efficient} propose learning strategies based on local exploration and global exploration to support the emergence of social norms.
Whereas their model focuses on maximizing the average payoffs among agents, \framework focuses on investigating the influence of social communication.
\citet{IJCAI-22:XSIGA} and \citet{IJCAI-18:Poros} show that robust norms emerge when agents deviating from norms provide rationales for their deviations, either in form of normative explanations \cite{IJCAI-22:XSIGA} or by revealing elements of their context \cite{IJCAI-18:Poros}. \framework complements these works---\framework agents learn norms via social communication from other agents observing deviations.

\citet{Morales2018offlineSynthesis} focus on the stability of synthesized norms that are verified by an evolutionary process.
\citet{Savarimuthu2010obligation} propose an algorithm to identify obligation norms based on association rule mining, a data mining technique.
\citet{Pernpeintner2021toward} proposes a governance approach that restricts action spaces based on publicly observable behaviors and transitions.

\citet{Levy2021convention} propose a framework that introduces congested actions where an agent's reward is not from pairwise interaction but is a function of others' actions and the environment.
\framework enables social learning from personal observation or by normative information sharing from explicit messages or softer notions of social communications including hints.

\revised{\citet{Sami2023reward} propose potential-based reward shaping via a convolutional neural network to construct the potential function on Hidden Markov Models while using message passing.
Whereas the messages in their work refer to the approximation of rewards in forward message and backward message, \framework regards messages and hints as the approach of agents for communicating normative information and acting as potential rewards.}

\section{Discussion}
\label{sec:discussion}

During and after COVID-19, abundant research has investigated the effects of interventions against the spread of the virus.
However, little research considers policy violations, which are the essential drivers of a pandemic.
Modeling social communications within our framework enables a more realistic simulation of individuals' decisions, e.g., obedience or noncompliance to interventions against the spread of the pandemic.

We present an approach that combines models of social communication to address the emergence of norms.
The novelty of our approach arises from its comprehensive treatment of the three main kinds of social communication that drive norm emergence: sanctions, tells, and hints.
Including normative information from tells and hints enables indirect social learning, which resembles human behaviors in the real world.

\subsection{Summary of Findings}

Our main findings are that agents who communicate via hints and respond to normative information converge to norms faster than those who respond only to hard sanctions or explicit communication of approval or disapproval.
Societies that consider hints are more robust in complying with the converged norms than those that do not consider hints. 
In our experiments, \framework and \EmoteSoc exceed the 90\% of norm emergence threshold faster than other societies and their compliance with the converged norm is higher than in \PrimitiveSoc, \SanctioningSoc, and \TellSoc.
\EmoteSoc does not defeat \SanctioningSoc and \TellSoc on all metrics but enhancing it to \framework makes a difference.

Specifically, in the pandemic environment simulation environment
(1) \framework enables better control on the spread of disease than other societies,
(2) agents in \framework and \EmoteSoc learn the self-isolation norm faster and are more willing to isolate themselves when infected,
(3) agents in \framework have higher goal satisfaction than the other societies.
As a result, \framework agents effectively avoid infection risk and yield higher satisfaction than baseline agents.

\subsection{Limitations and Threats to Validity}

We identify and mitigate three threats.
First, humans have limited observations of each other and may deceive or withhold information.
To mitigate this threat in evaluation,
we made simplifying assumptions that agents can infer each other's communications and that all communication between agents is genuine and honest.
Second, we base our simulation on predefined payoffs and probabilities.
Obtaining accurate valuations of disease transition may require significant time and effort.
We mitigate this threat by apply data from the literature.
Third, variations exist in the severity of penalties and the probabilities of softer sanctions within social communications.
Our study presents societies with diverse forms of social communications to demonstrate the efficacy of our approach.

\subsection{Future Directions}

As AI becomes part of our daily lives, incorporating human ethics into AI becomes a necessary problem \cite{AIES-18:ethics,Lopez2017norms,Computer-23:Wasabi,Woodgate+Ajmeri-22:ethics-STS}.
Since human behavior is driven by the pursuit of values, studying human values helps us understand human decisions and create agents that reason over human values \cite{Ajmeri2020elessar,Murukannaiah2020Blue-Sky,Liscio2021axies}. Social communication could also convey values.
Whereas \citet{Montes2021value} automate norm synthesis based on value promotion, an interesting direction is to embed values into autonomous agents.
That is, how can we develop agents that are capable of making value-aligned decisions?
A line of future research is to investigate dimensions of emotions, physical arousal, that describes the strength of the emotional state \cite{Collins+COINE23-Svoie}.
Another future direction includes considering a mix of personality types in \framework \cite{COINE-22:Fleur}.
We can investigate how different values influence human interactions in future research to support high heterogeneity.

\begin{acks}
We thank the anonymous reviewers for helpful comments.
ST and MPS thank the NSF (grant IIS-2116751) for partial support for this research.
\end{acks}

\bibliographystyle{ACM-Reference-Format}
\bibliography{Munindar,Nirav,SzTing}

\clearpage
\appendix

\section{Code and Parameters}
\label{sec:reproducibility}

Table~\ref{tbl:parameters} lists the hyperparameters in our simulation.
The codebase for our social simulation is available on GitHub: \url{https://github.com/ChristineTzeng/Nest-AAMAS2024/}.

\begin{table*}[!htb]
\centering
\caption{Hyperparameters.}
\label{tbl:parameters}

\begin{tabular}{p{4cm} n{3}{3} p{8cm}}\toprule
Parameter & \multicolumn{1}{r}{Value} & Comment \\\midrule
Learning rate $\alpha$ & 0.001 & The parameter that controls the rate of update on the estimation of state-action values \\
Discount factor $\gamma$ & 0.9 & The parameter that defines if future rewards would be considered \\
Simulation step per action & 1 & The duration of each action's timesteps \\
Population size & 100 & The number of agents in one society \\
Infection \% & 0.3 & The default fraction of infected agents in a society\\
Certainty of potential reward & 0.3 & Value for $\kappa$ for certainty of possible sanctions from normative information through hints \\
Certainty of potential reward & 0.5 & Value for $\kappa$ for certainty of possible sanctions from normative information messages\\
\bottomrule
\end{tabular}
\end{table*}



\section{Additional Results}
\label{sec:additional-results}

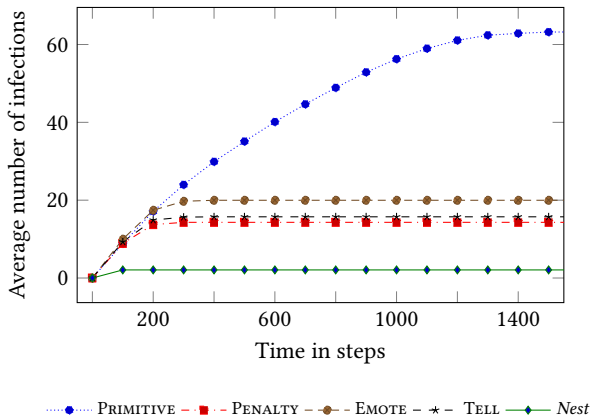
\begin{figure}[!htb]
\centering
    \begin{tikzpicture}
        \begin{axis}[
        title={},
        height=0.65\columnwidth,
        width=0.95\columnwidth,
        xlabel={Time in steps},
        ylabel={Average number of infections},
        xticklabels={,,200,,600,,1000,,1400},
        xmin=-50,xmax=1550,
        legend style={at={(0.5,-0.3)},anchor=north,},
        legend columns=5,
        legend style={fill=none,draw=none}
        ]
        \addplot +[mark size=1.5, each nth point={100},  densely dotted] table [x=steps, y=total_number_infections, col sep=comma]
        {data/Ness/primitive_rolling_50.csv};
        \addplot +[mark size=1.5, each nth point={100},  dashdotted] table [x=steps, y=total_number_infections, col sep=comma]
        {data/Ness/sanctioning_rolling_50.csv};
        \addplot +[mark size=1.5, each nth point={100},  densely dashed] table [x=steps, y=total_number_infections, col sep=comma]
        {data/Ness/feeling_probE0.5_rolling_50.csv};
        \addplot +[mark size=1.5, each nth point={100},  dashed] table [x=steps, y=total_number_infections, col sep=comma]
        {data/Ness/message_prob_rolling_50.csv};
        \addplot +[green!50!black, mark color=green, mark size=1.5, each nth point={100}] table [x=steps, y=total_number_infections, col sep=comma]
        {data/Ness/hint_probE0.5_rolling_50.csv};
        \legend{\footnotesize \PrimitiveSoc, \footnotesize \SanctioningSoc, \footnotesize \EmoteSoc, \footnotesize \TellSoc, \footnotesize \framework}
        \end{axis}
    \end{tikzpicture}

\caption{
Comparing the average number of infections (M\fsub{Infections}) in various societies.
\framework yields fewer infections on average than other societies.
The effect is large.
}
\label{fig:tot_infections-supplement}
\end{figure}

Figure~\ref{fig:tot_infections-supplement} plots the total number of infected agents in various societies.

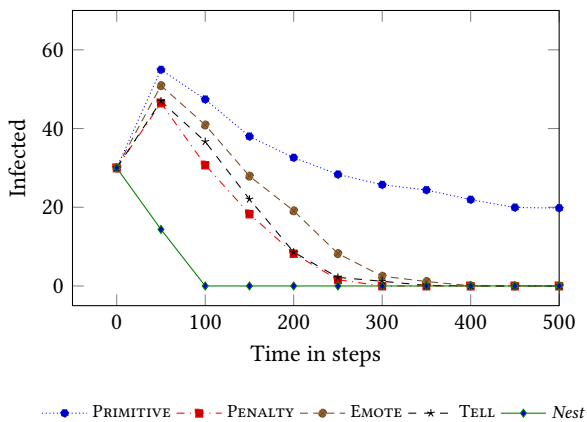
\begin{figure}[!htb]
\centering
    \begin{tikzpicture}
    \begin{axis}[
    title={},
    height=0.65\columnwidth,
    width=0.95\columnwidth,
    xlabel={Time in steps},
    ylabel={Infected},
    xmin=-50,xmax=500,
    ymin=-5,ymax=70,
    legend style={at={(0.5,-0.3)},anchor=north,},
    legend columns=5,
    legend style={fill=none,draw=none}
    ]
    \addplot +[mark size=1.5, each nth point={50},  densely dotted] table [x=steps, y=infected, col sep=comma]
    {data/Ness/primitive_rolling_50.csv};
    \addplot +[mark size=1.5, each nth point={50},  dashdotted] table [x=steps, y=infected, col sep=comma]
    {data/Ness/sanctioning_rolling_50.csv};
    \addplot +[mark size=1.5, each nth point={50},  densely dashed] table [x=steps, y=infected, col sep=comma]
    {data/Ness/feeling_probE0.5_rolling_50.csv};
    \addplot +[mark size=1.5, each nth point={50},  dashed] table [x=steps, y=infected, col sep=comma]
    {data/Ness/message_prob_rolling_50.csv};
    \addplot +[green!50!black, mark color=green, mark size=1.5, each nth point={50}] table [x=steps, y=infected, col sep=comma]
    {data/Ness/hint_probE0.5_rolling_50.csv};
    \legend{\footnotesize \PrimitiveSoc, \footnotesize \SanctioningSoc, \footnotesize \EmoteSoc, \footnotesize \TellSoc, \footnotesize \framework}
    \end{axis}
    \end{tikzpicture}
\caption{
Comparing the number of infected agents in the first 500 steps. The differences are noticeable here.
}
\label{fig:infected-500}
\end{figure}

\begin{figure}[!htb]
\centering
    \begin{tikzpicture}
    \begin{axis}[
    title={},
    height=0.65\columnwidth,
    width=0.95\columnwidth,
    xlabel={Time in steps},
    ylabel={Deceased},
    xmin=-50,xmax=500,
    legend style={at={(0.5,-0.3)},anchor=north,},
    legend columns=5,
    legend style={fill=none,draw=none}
    ]
    \addplot +[mark size=1.5, each nth point={50},  densely dotted] table [x=steps, y=deceased, col sep=comma]
    {data/Ness/primitive_rolling_50.csv};
    \addplot +[mark size=1.5, each nth point={50},  dashdotted] table [x=steps, y=deceased, col sep=comma]
    {data/Ness/sanctioning_rolling_50.csv};
    \addplot +[mark size=1.5, each nth point={50},  densely dashed] table [x=steps, y=deceased, col sep=comma]
    {data/Ness/feeling_probE0.5_rolling_50.csv};
    \addplot +[mark size=1.5, each nth point={50},  dashed] table [x=steps, y=deceased, col sep=comma]
    {data/Ness/message_prob_rolling_50.csv};
    \addplot +[green!50!black, mark color=green, mark size=1.5, each nth point={50}] table [x=steps, y=deceased, col sep=comma]
    {data/Ness/hint_probE0.5_rolling_50.csv};
    \legend{\footnotesize \PrimitiveSoc, \footnotesize \SanctioningSoc, \footnotesize\EmoteSoc, \footnotesize \TellSoc, \footnotesize \framework}
    \end{axis}
    \end{tikzpicture}
\caption{
Comparing the number of deceased agents in the first 500 steps. The differences between \framework and \EmoteSoc are noticeable here.
}
\label{fig:deceased-500}
\end{figure}
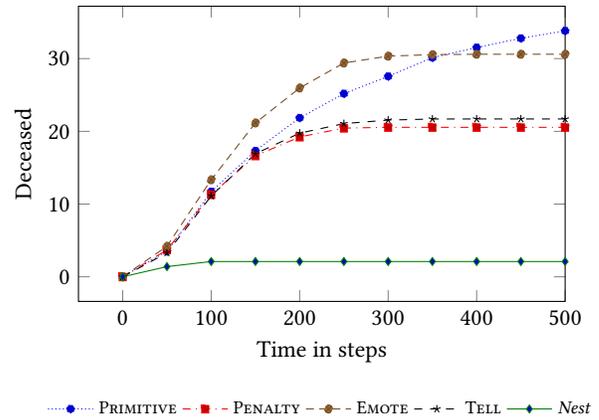

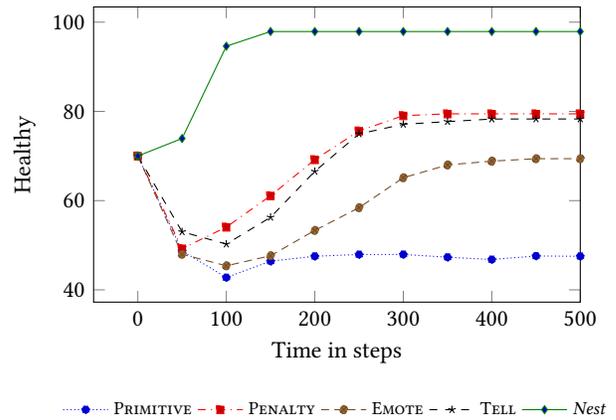
\begin{figure}[!htb]
\centering
    \begin{tikzpicture}
    \begin{axis}[
    title={},
    height=0.65\columnwidth,
    width=0.95\columnwidth,
    xlabel={Time in steps},
    ylabel={Healthy},
    xmin=-50,xmax=500,
    legend style={at={(0.5,-0.3)},anchor=north,},
    legend columns=5,
    legend style={fill=none,draw=none}
    ]
    \addplot +[mark size=1.5, each nth point={50},  densely dotted] table [x=steps, y=healthy, col sep=comma]
    {data/Ness/primitive_rolling_50.csv};
    \addplot +[mark size=1.5, each nth point={50},  dashdotted] table [x=steps, y=healthy, col sep=comma]
    {data/Ness/sanctioning_rolling_50.csv};
    \addplot +[mark size=1.5, each nth point={50},  densely dashed] table [x=steps, y=healthy, col sep=comma]
    {data/Ness/feeling_probE0.5_rolling_50.csv};
    \addplot +[mark size=1.5, each nth point={50},  dashed] table [x=steps, y=healthy, col sep=comma]
    {data/Ness/message_prob_rolling_50.csv};
    \addplot +[green!50!black, mark color=green, mark size=1.5, each nth point={50}] table [x=steps, y=healthy, col sep=comma]
    {data/Ness/hint_probE0.5_rolling_50.csv};
    \legend{\footnotesize \PrimitiveSoc, \footnotesize \SanctioningSoc, \footnotesize \EmoteSoc, \footnotesize \TellSoc, \footnotesize \framework}
    \end{axis}
    \end{tikzpicture}
\caption{
Comparing the number of healthy agents in the first 500 steps. The differences are noticeable here.
}
\label{fig:healthy-500}
\end{figure}

\begin{figure}[!htb]
\centering
    \begin{tikzpicture}
    \begin{axis}[
    title={},
    height=0.65\columnwidth,
    width=0.95\columnwidth,
    xlabel={Time in steps},
    ylabel={Vaccinated},
    xmin=-50,xmax=500,
    legend style={at={(0.5,-0.3)},anchor=north,},
    legend columns=5,
    legend style={fill=none,draw=none}
    ]
    \addplot +[mark size=1.5, each nth point={50},  densely dotted] table [x=steps, y=vaccinated, col sep=comma]
    {data/Ness/primitive_rolling_50.csv};
    \addplot +[mark size=1.5, each nth point={50},  dashdotted] table [x=steps, y=vaccinated, col sep=comma]
    {data/Ness/sanctioning_rolling_50.csv};
    \addplot +[mark size=1.5, each nth point={50},  densely dashed] table [x=steps, y=vaccinated, col sep=comma]
    {data/Ness/feeling_probE0.5_rolling_50.csv};
    \addplot +[mark size=1.5, each nth point={50},  dashed] table [x=steps, y=vaccinated, col sep=comma]
    {data/Ness/message_prob_rolling_50.csv};
    \addplot +[green!50!black, mark color=green, mark size=1.5, each nth point={50}] table [x=steps, y=vaccinated, col sep=comma]
    {data/Ness/hint_probE0.5_rolling_50.csv};
    \legend{\footnotesize \PrimitiveSoc, \footnotesize \SanctioningSoc, \footnotesize \EmoteSoc, \footnotesize \TellSoc, \footnotesize \framework}
    \end{axis}
    \end{tikzpicture}
\caption{
Comparing the number of vaccinated agents in the first 500 steps. The differences between \framework and other societies are noticeable here.
}
\label{fig:vaccinated-500}
\end{figure}
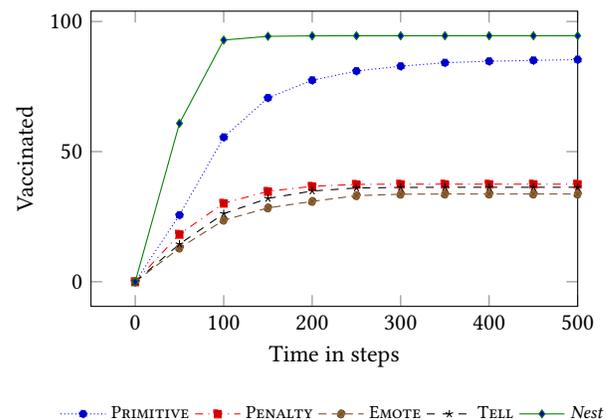

Figures~\ref{fig:infected-500}, \ref{fig:deceased-500}, \ref{fig:healthy-500}, and \ref{fig:vaccinated-500} shows plots for the number of infected, deceased, healthy, and vaccinated agents in the first 500 steps in various societies, where the differences are noticeable.

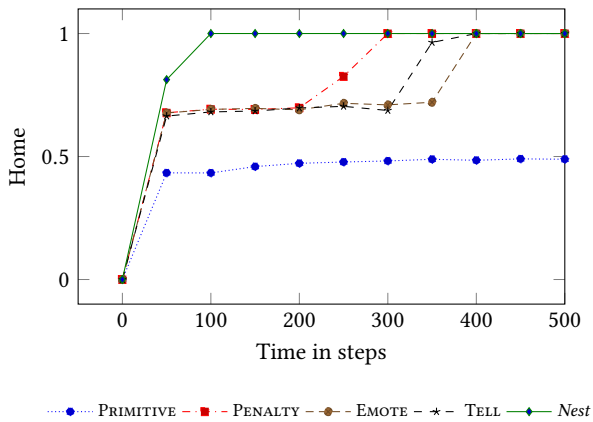
\begin{figure}[!htb]
\centering

\begin{tikzpicture}
    \begin{axis}[
    title={},
    height=0.65\columnwidth,
    width=0.95\columnwidth,
    xlabel={Time in steps},
    ylabel={Home},
    xmin=-50,xmax=500,
    legend style={at={(0.5,-0.3)},anchor=north,},
    legend columns=5,
    legend style={fill=none,draw=none}
    ]
    \addplot +[mark size=1.5, each nth point={50},  densely dotted] table [x=steps, y=isolation, col sep=comma]
    {data/Ness/primitive_rolling_50.csv};
    \addplot +[mark size=1.5, each nth point={50},  dashdotted] table [x=steps, y=isolation, col sep=comma]
    {data/Ness/sanctioning_rolling_50.csv};
    \addplot +[mark size=1.5, each nth point={50},  densely dashed] table [x=steps, y=isolation, col sep=comma]
    {data/Ness/feeling_probE0.5_rolling_50.csv};
    \addplot +[mark size=1.5, each nth point={50},  dashed] table [x=steps, y=isolation, col sep=comma]
    {data/Ness/message_prob_rolling_50.csv};
    \addplot +[green!50!black, mark color=green, mark size=1.5, each nth point={50}] table [x=steps, y=isolation, col sep=comma]
    {data/Ness/hint_probE0.5_rolling_50.csv};
    \legend{\footnotesize \PrimitiveSoc, \footnotesize \SanctioningSoc, \footnotesize \EmoteSoc, \footnotesize \TellSoc, \footnotesize \framework}
    \end{axis}
    \end{tikzpicture}
\caption{
Comparing the number of agents in isolation (M\fsub{Home}) in \framework and other societies in the first 500 steps.
}
\label{fig:compliance-500-home}
\end{figure}

\begin{figure}[!htb]
\centering
    \begin{tikzpicture}
    \begin{axis}[
    title={},
    height=0.65\columnwidth,
    width=0.95\columnwidth,
    xlabel={Time in steps},
    ylabel={Quarantine},
    xmin=-50,xmax=500,
    legend style={at={(0.5,-0.3)},anchor=north,},
    legend columns=4,
    legend style={fill=none,draw=none}
    ]
    \addplot +[mark size=1.5, each nth point={50}, densely dotted] table [x=steps, y=forced_quarantine, col sep=comma]
    {data/Ness/primitive_rolling_50.csv};
    \addplot +[mark size=1.5, each nth point={50}, dashdotted] table [x=steps, y=forced_quarantine, col sep=comma]
    {data/Ness/sanctioning_rolling_50.csv};
    \addplot +[mark size=1.5, each nth point={50}, densely dashed] table [x=steps, y=forced_quarantine, col sep=comma]
    {data/Ness/feeling_probE0.5_rolling_50.csv};
    \addplot +[mark size=1.5, each nth point={50}, dashed] table [x=steps, y=forced_quarantine, col sep=comma]
    {data/Ness/message_prob_rolling_50.csv};
    \addplot +[green!50!black, mark color=green, mark size=1.5, each nth point={50}] table [x=steps, y=forced_quarantine, col sep=comma]
    {data/Ness/hint_probE0.5_rolling_50.csv};
    \legend{, \footnotesize \SanctioningSoc, \footnotesize \EmoteSoc, \footnotesize \TellSoc, \footnotesize \framework}
    \end{axis}
    \end{tikzpicture}

\caption{
Comparing the number of agents in quarantine (M\fsub{Quarantine}) in \framework and other societies in the first 500 steps.
}
\label{fig:compliance-500-quarantine}
\end{figure}
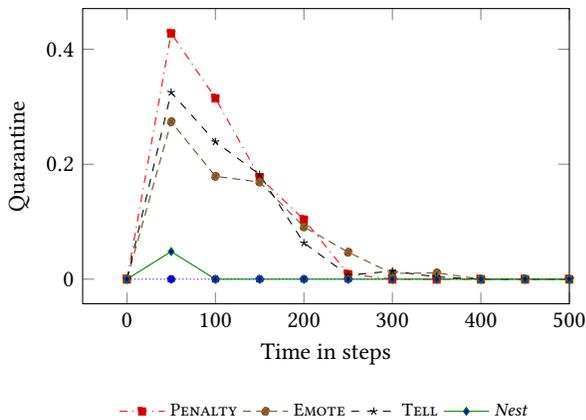

Figure~\ref{fig:compliance-500-home} and Figure~\ref{fig:compliance-500-quarantine} show the number of agents in isolation and quarantine, respectively, in various societies in the first 500 steps of the simulation.

\clearpage

\end{document}